\documentclass[prl,reprint,twocolumn,footinbib, amsmath,amssymb, aps,superscriptaddress]{revtex4}

\usepackage{graphicx}
\usepackage{color}
\usepackage{amsmath}
\usepackage{booktabs}
\usepackage{float}
\begin{document}
\preprint{APS/123-QED}

\title{Tunable Shear Thickening in Suspensions}
\author{Neil Y.C. Lin}
\affiliation{Department of Physics, Cornell University, Ithaca, New York 14853}
\author{Christopher Ness}
\affiliation{School of Engineering, University of Edinburgh, Edinburgh EH9 3JL, United Kingdom}
\author{Michael E. Cates}
\affiliation{DAMTP, University of Cambridge, Centre for Mathematical Sciences, Wilberforce Road, Cambridge CB3 0WA, United Kingdom}
\author{Jin Sun}
\affiliation{School of Engineering, University of Edinburgh, Edinburgh EH9 3JL, United Kingdom}
\author{Itai Cohen}
\affiliation{Department of Physics, Cornell University, Ithaca, New York 14853}
\date{\today}

\begin{abstract}
Shear thickening, an increase of viscosity with shear rate, is a ubiquitous phenomena in suspended materials that has implications for broad technological applications. Controlling this thickening behavior remains a major challenge and has led to empirical strategies ranging from altering the particle surfaces and shape to modifying the solvent properties. However, none of these methods allow for tuning of flow properties during shear itself. Here, we demonstrate that by strategic imposition of a high-frequency and low-amplitude shear perturbation orthogonal to the primary shearing flow, we can largely eradicate shear thickening. The orthogonal shear effectively becomes a regulator for controlling thickening in the suspension, allowing the viscosity to be reduced by up to two decades on demand. In a separate setup, we show that such effects can be induced by simply agitating the sample transversely to the primary shear direction. Overall, the ability of in situ manipulation of shear thickening paves a route towards creating materials whose mechanical properties can be controlled.
\end{abstract}

\maketitle

The viscosity of a densely packed suspension of particles can increase radically when sheared beyond a critical stress~\cite{wagner2009shear,guy2015towards}. This thickening behavior has been exploited in technological applications ranging from vehicle traction control to flexible spacesuits that protect astronauts from micrometeorite impacts~\cite{wagner2007advanced, lee2003ballistic,cwalina2016shear}. It may also lead to flow problems such as pipe blockage during industrial extrusion processes~\cite{brown2014shear}. Shear thickening has generally been considered an inherent material property~\cite{Brown2010}, rather than as a response that can be tuned. As a consequence, suspension process design is often constrained within tight bounds to avoid thickening ~\cite{Benbow1993}, while the applications of such flow behavior are limited by a lack of tunability.

To design our control method, we take advantage of the underlying shear thickening mechanisms that have been revealed recently. Experiments and simulations have shown that when the stress applied to a suspension of micron-sized particles exceeds a critical value, the particle-particle interaction switches from lubricated to frictional, enhancing resistance to flow~\cite{Fernandez2013, seto2013discontinuous, wyart2014discontinuous, lin2015hydrodynamic, brown2014shear, brown2012role}. The stress is transmitted through shear-induced force chains, which arise from frictional particle contacts~\cite{bi2011jamming, mari2014shear, brown2014shear, brown2012role}, aligned along the compressive axis. Such chains are fragile~\cite{cates1998jamming, majmudar2005contact} and are constantly broken and rebuilt during steady shear.

\begin{figure*} [htp]
\centering
\includegraphics[width=0.7 \textwidth]{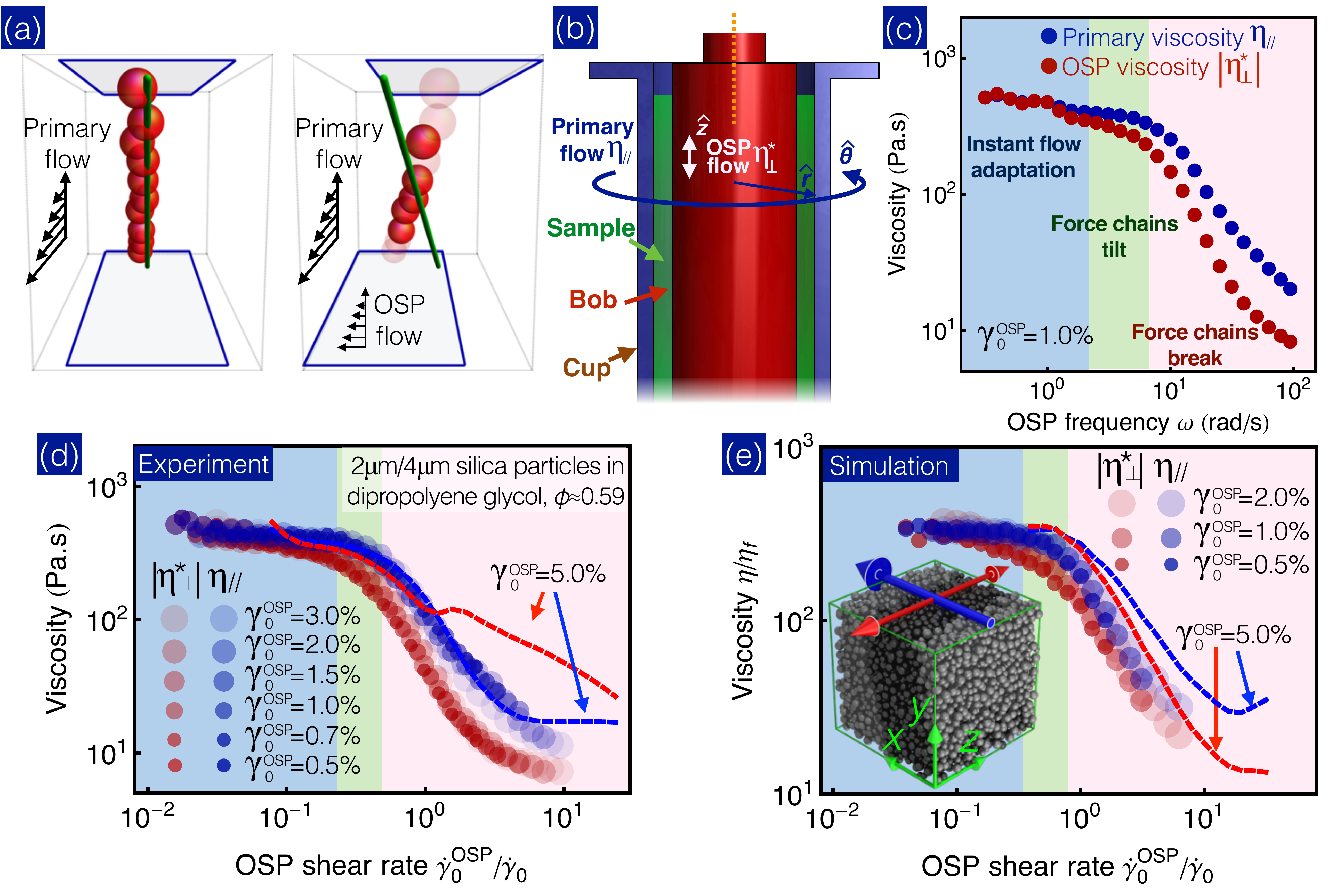}
\caption{Perturbing the frictional force chains that govern shear thickening by imposing orthogonal shear can reduce the suspension viscosity by nearly two decades. (a) Schematic showing force chain alignment during flow. Left panel: Under simple, uniaxial shear, particles (red spheres) naturally align along the compressive axis (green line); Right panel: Under rapid biaxial shear, with transverse flow indicated by the shifted upper plate, the orthogonal perturbations may tilt or break the chains, so that they deviate from the combined compressive axis. (b) Schematic of experimental apparatus showing the continuous primary flow that drives shear thickening and the inner oscillatory module that perturbs the force chains; (c) Frequency sweep experimental data for $\gamma_0^\mathrm{OSP}$=1\%. Shown are the primary viscosity ($\eta_\parallel$, blue dots) and orthogonal complex viscosity ($|\eta_\bot^*|$, red dots); (d) Experimental and (e) simulation viscosity data collapse as function of the relative OSP shear rate ($\dot \gamma^\mathrm{OSP}_0/\dot \gamma_0$), for several $\gamma_0^\mathrm{OSP}$. Dashed lines indicate reentrant thickening when the OSP strain becomes large enough to induce new force chains; Inset (e), snapshot of simulation configuration, indicating the coordinate definitions with respect to the primary (blue arrow) and OSP (red arrows) flow directions. Further details and characterization of experimental sample can be found in Method and SI.}
\label{figure1}
\end{figure*}

This fragility paradigm asserts that these stress-transmitting chains are themselves a product of the stress, with a finite chain-assembly time required following startup or perturbations to the flow direction~\cite{gadala1980shear, ness2016two}. These insights suggest a strategy for controlling thickening. For perturbations slower than chain assembly, contact rearrangement is sufficiently fast that force chains remain aligned with the instantaneous net compressive axis. Conversely, for perturbations faster than the assembly time, chains cannot reach compatibility with the instantaneous net compressive axis, but occupy a partially-assembled transient state, illustrated in Fig.~\ref{figure1}(a). The alignment of the perturbed or tilted chain deviates from the net compressive axis and it no longer transmits stresses effectively. Thus it should be possible to precisely regulate the thickening behavior by applying appropriate lateral perturbations. 

To that end we design a \textit{biaxial} shear protocol that uses an orthogonal flow perturbation to interfere with force chains induced by a primary shearing flow. Our strategy is to maximize the perturbation influence so the force chains usually responsible for thickening cannot establish fully. We conduct biaxial rheometry experimentally and numerically, mapping the response of a hard-sphere suspension as the perturbation rate and amplitude are systematically varied.
By integrating our knowledge of the force chain alignment, mechanical instability and direct link to the viscosity, we show how this strategy can be optimized.  
We focus on \textit{discontinuous} shear thickening suspensions, as their vast viscosity variations make them most problematic to the engineer~\cite{wyart2014discontinuous, seto2013discontinuous, fall2008shear}. Our results show that through suitable regulation, the suspension viscosity at a fixed flow rate may be reduced by up to two decades in an active and controlled manner. We finally demonstrate the wide utility of the technique using a simpler flow regulation set up.

The biaxial rheometry experiment is performed using a double-wall Couette geometry that has an outer cup driven continuously by an underneath motor, and an inner bob attached to an oscillating shaft~\cite{vermant1998orthogonal, vermant1997orthogonal, jacob2015convective}. A simplified schematic and coordinate definition are shown in Fig.~\ref{figure1}(b). The continuous primary flow in the $\hat r \hat\theta$ plane along the $\hat\theta$ direction, at controlled rate $\dot{\gamma}_0$, constantly induces force chains, sets the initial shear thickening state, and probes the parallel viscosity, $\eta_\parallel$. Concurrently, an orthogonal superimposed perturbation (OSP) comprising oscillatory flow in the $\hat r \hat z$ plane along the $\pm \hat z$ direction perturbs the suspension by imposing a deformation ${\gamma}^\mathrm{OSP} = \gamma_0^\mathrm{OSP}\sin(\omega t)$ with rate $\dot{\gamma}^\mathrm{OSP} = \omega\gamma_0^\mathrm{OSP}\cos(\omega t)$, and simultaneously probes its orthogonal complex viscosity, $|\eta_\bot^*|$. Given the employed Couette cell dimension, we approximate the primary flow as a uniaxial shear in parallel plate geometry\footnote[1]{Note that uniaxial \textit{shear} as defined here is technically a biaxial \textit{flow} since there is no axis of rotational symmetry}. Further details about the geometry and OSP calibration can be found in Method and SI.

Following the above reasoning, we take the primary shear timescale $1/\dot{\gamma}_0$ and the OSP period $1/\omega$ as proxies for chain-assembly and perturbation times, respectively. We fix the primary shear rate at $\dot \gamma_0 = 0.2$ s$^{-1}$, where the suspension is normally strongly thickened (see SI), and conduct an OSP frequency sweep at a fixed strain amplitude $\gamma^\mathrm{OSP}_0=1\%$. The evolutions of $\eta_\parallel$ and $|\eta_\bot^*|$ are given in Fig.~\ref{figure1}(c).

At low frequencies $\omega\le 2$ rad/s (blue area, Fig.~\ref{figure1}(c)), $\eta_\parallel$ and $|\eta_\bot^*|$ exhibit plateaus that match the original thickening viscosity $\sim 500$ Pa.s, for $\gamma_0^\mathrm{OSP} = 0$ (see SI). In this regime, the primary flow renews the force chains rapidly so that they are always compatible with the net compressive axis, and their alignment appears to adapt instantly to the OSP flow. This instantaneous adaptation gives rise to an isotropic and unaffected thickening behaviour (see SI for an analytical derivation). As the frequency increases to  2 rad/s $< \omega\le 8$ rad/s (green area), we find that the OSP viscosity $|\eta_\bot^*|$ decays slightly, while the primary value $\eta_\parallel$ remains relatively constant (within 7\%). The decay in $|\eta_\bot^*|$ suggests that the OSP flow deforms or tilts existing force chains more quickly than they are replaced by new ones, while the unchanged $\eta_\parallel$ suggests the tilted chains remain largely intact. Finally, we observe a substantial drop in both  $\eta_\parallel$ and $|\eta_\bot^*|$ at even higher frequencies $\omega > 8$ rad/s implying significant breakage of force chains and dissolution of the flow induced contact network. Qualitatively, this result is consistent with our above interpretation of the fragile, marginally stable nature of the force chain network, and reaffirms the link between the stress bearing capability of the chains, and the viscosity of the suspension~\cite{wyart2014discontinuous}.

\begin{figure*} [htp]
\centering
\includegraphics[width=0.75 \textwidth]{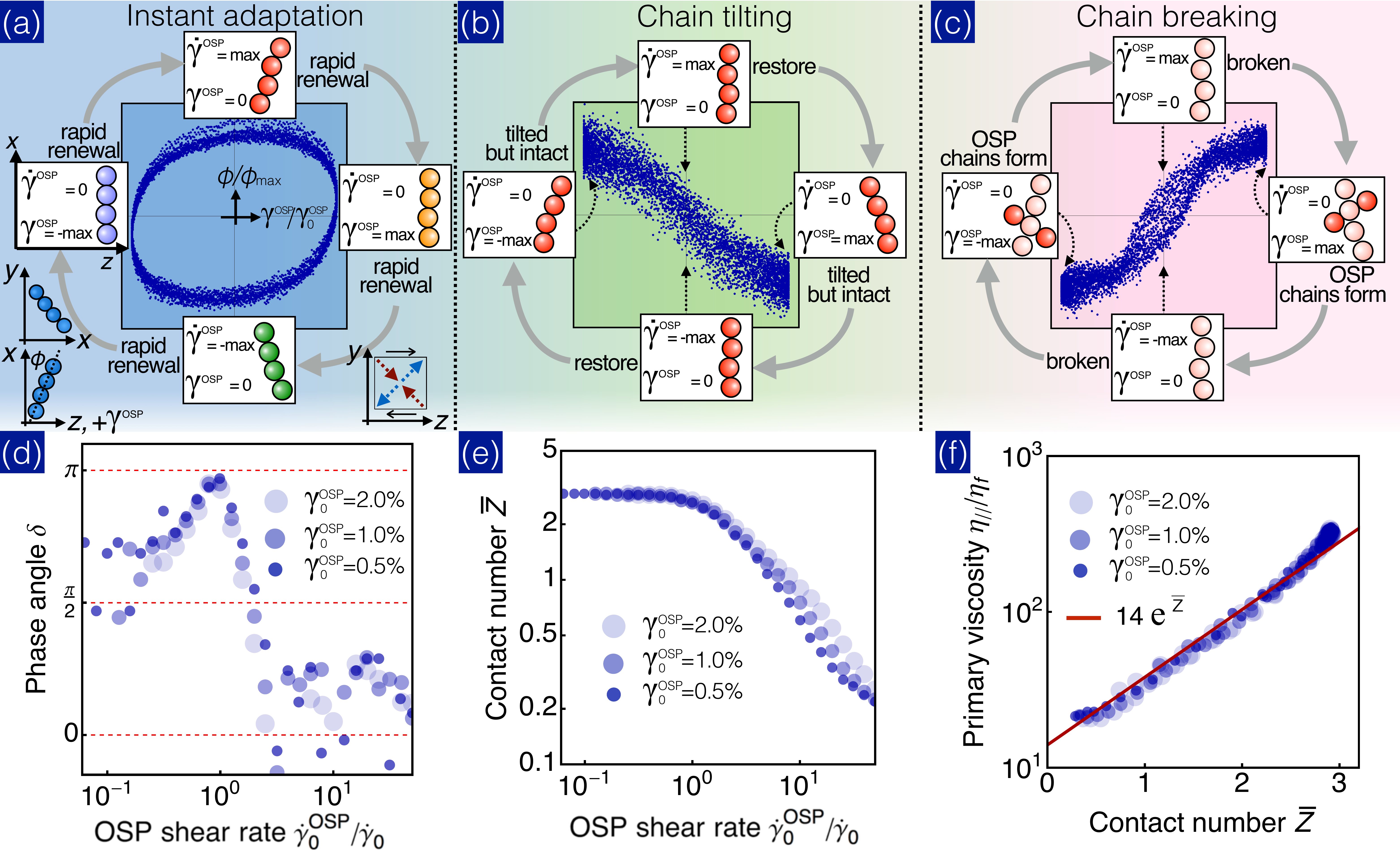}
\caption{Microstructural information from simulations to illustrate the mechanism behind the viscosity reduction with increased OSP flow, for $\gamma_0^\mathrm{OSP}=1\%$. Shown are Lissajous curves for microstructural force chain alignment for the (a) instant adaptation, (b) chain tilting and (c) chain breaking regimes, with $\phi$ and $\gamma^\mathrm{OSP}$ normalized by their maximal values, where $\phi_{\mathrm{max}}\sim \mathcal{O}(10^{-2})$. $\phi$ and the sign of $\gamma^\mathrm{OSP}$ are defined in the Left Inset of (a), and definitions of compression (red) and extension (blue) quadrants of the OSP flow in Right Inset of (a), for positive $\dot{\gamma}^\mathrm{OSP}$. Force chain diagrams in (a), (b), (c) illustrate alignment projected to $\hat x \hat z$ at each stage of the oscillatory cycle, while alignment in $\hat x \hat y$ remains along the compressive axis of the primary flow throughout (Left Inset of (a)); (d) Evolution of phase angle $\delta$ between $\phi$ and $\gamma^\mathrm{OSP}$ with OSP shear rate, where $\delta = 0$ and $\delta=\pi/2$ respresent elastic and viscous responses respectively; (e) Evolution of time-averaged contact number $\bar{Z}$ with OSP shear rate; (f) Direct dependence of primary flow viscosity on time-averaged contact number $\bar{Z}$.}
\label{figure2}
\end{figure*}

To further elucidate the roles being played by the primary and OSP flows, we repeat the frequency sweep measurement at six different $\gamma^\mathrm{OSP}_0$, Fig.~\ref{figure1}(d). We normalize the OSP shear rate magnitude by the primary rate $\dot \gamma^\mathrm{OSP}_0/\dot \gamma_0 (= \omega \gamma^\mathrm{OSP}_0 / \dot \gamma_0$), and find that all $\eta_\parallel$ and $|\eta_\bot^*|$ data, for $\gamma^\mathrm{OSP}_0<$ 5\%, collapse onto two master curves (blue and red dots, respectively). Furthermore, the onset of the chain breaking regime occurs at $\dot \gamma^\mathrm{OSP}_0/\dot \gamma_0 \approx 1$. This scaling suggests that the force chain response, whether they instantly adapt to the OSP flow, tilt, or break, is determined by the competition between $\dot \gamma^\mathrm{OSP}_0$ and $\dot \gamma_0$. While $\dot \gamma_0$ indicates how frequently force chains reform, $\dot \gamma^\mathrm{OSP}_0$ dictates how rapidly the OSP flow perturbs these structures. In other words, the ratio $\dot{\gamma}_0^\mathrm{OSP}/\dot{\gamma}_0$ directly governs the viscosity reduction of a biaxially sheared sample. For large $\gamma_0^\mathrm{OSP}$ ($>$5\%), we observe a deviation from the master curves, suggesting a possible reentrant thickening arising when chains induced by the OSP flow emerge. We conclude, therefore, that for sufficiently large $\dot{\gamma_0}^\mathrm{OSP}/\dot{\gamma}_0$ and sufficiently small $\gamma_0^\mathrm{OSP}$, our orthogonal flow perturbation disrupts the conventional shear-induced contact network, inhibiting friction-mediated force chains and mitigating shear thickening. The extent of this mitigation may be set using $\dot \gamma^\mathrm{OSP}_0/\dot \gamma_0$ as a control parameter, allowing precise \emph{regulation} of the viscosity of dense suspensions.

To clarify the detailed microstructural rearrangements leading to the observed measurements we use numerical simulations, explicitly resolving the trajectories and interactions of suspended, bidisperse spheres (diameter ratio 1:1.4) using a classical discrete element code~\cite{Cundall1979,plimpton1995fast} enhanced with the recently established Critical Load Model~\cite{mari2014shear,ness2016shear}. Hydrodynamic forces are approximated as pairwise lubrication interactions (viscosity $\eta_f$) between neighboring particles, while particle contacts are treated as linear springs with friction appearing above a critical normal force. The fixed-volume fraction (55\%), periodic, Cartesian simulation cell can be deformed to simulate steady shear in a primary direction with a small amplitude oscillation in the orthogonal direction. With respect to the illustration in Fig~\ref{figure1}(e), primary flow (blue arrow) is in the $\hat x \hat y$ plane, along the $\hat x$ direction, and OSP flow (red arrows) is in the $\hat y \hat z$ plane, along the $\pm \hat z$ direction. Full details and discussion of model assumptions are given in SI.

The experimental flow measurements are repeated computationally with consistent results, shown in Fig.~\ref{figure1}(e). The force chain alignment is interrogated in the three identified regimes, instant adaptation, chain tilting, and chain breaking. In all regimes, the chains lie mainly along the compressive axis of the primary shear flow, but are subtly shifted out of the $\hat x \hat y$ plane by the OSP flow. To quantify these deviations, we calculate a fabric tensor $\langle \hat r^{\alpha \beta}_i \hat r^{\alpha \beta}_j \rangle$ capturing the particle contact configuration~\cite{bi2011jamming}, where $r^{\alpha \beta}_i$ ($\alpha \neq \beta$) is the unit vector between particles $\alpha$ and $\beta$, while $i, j$ denote the coordinate indices and $\langle \ldots \rangle$ denotes the average over all neighboring particles. We take $\phi = \tan^{-1}( \langle \hat r_y \hat r_z  \rangle / \langle \hat r_x \hat r_y  \rangle)$ as the angle between the force chains and $x$-axis, when projected to $\hat x \hat z$, Fig~\ref{figure2}a (Left Inset), which remains small, $\phi_\mathrm{max} \sim \mathcal{O}(10^{-2})$, throughout. Chains inclined towards the compressive quadrant of the OSP flow (see Right Inset, Fig~\ref{figure2}a) have positive $\phi$ for positive $\dot{\gamma}^\mathrm{OSP}$.

Representative Lissajous curves of $\phi(\gamma^\mathrm{OSP})$ for the three regimes are shown in Fig.~\ref{figure2}(a), (b), and (c), for time-varying OSP strain $\gamma^\mathrm{OSP}$.
For low OSP rates $\dot \gamma^\mathrm{OSP}_0/\dot{\gamma}_0\ll 1$, newly formed force chains always align with the net compressive axis dictated by the instantaneous combined flow, so $\phi$ is in phase with the time-varying OSP \textit{strain rate}, $\phi \propto \dot{\gamma}^\mathrm{OSP}$, Fig.~\ref{figure2}(a), generating isotropic thickening as reported in Fig.~\ref{figure1}(d)~and~(e).
For $\dot \gamma^\mathrm{OSP}_0/\dot{\gamma}_0 \approx 1$, the time scales for the OSP and primary flows are comparable, so it is expected that chains are tilted before they can rearrange.
Indeed, we find that chain alignment is antiphase to the OSP \textit{strain}, $\phi \propto -\gamma^\mathrm{OSP}$, Fig.~\ref{figure2}(b).
This indicates that rather than adapting to the instantaneous flow rate, existing force chains are instead tilted affinely by the transverse deformation (strain) and occupy the extensional quadrant of the OSP flow while $\gamma^\mathrm{OSP}\neq0$, as sketched in Fig~\ref{figure1}a (right panel). 
When the OSP rate dominates $\dot \gamma^\mathrm{OSP}_0/\dot{\gamma}_0 \gg 1$, the affine deformation caused by the transverse flow is fast enough to break these tilted force chains more rapidly than the primary flow is able to sustain or re-establish them.
Meanwhile, new contacts may be formed by the OSP flow in its compressive quadrant. Thus, the chain alignment is in phase with the OSP strain $\phi\propto \gamma^\mathrm{OSP}$, Fig.~\ref{figure2}(c).
The implication of the final curve is striking. In the chain breaking regime, the thickening becomes solely governed by the strain amplitude of the OSP flow. We summarize the chain alignment behavior by plotting the phase lag $\delta$ between $\phi$ and $\gamma^\mathrm{OSP}$ in Fig.~\ref{figure2}(d).

Since the shear thickened viscosity directly arises from frictional particle contacts~\cite{lin2015hydrodynamic}, we expect that the evolution of the force chain response with increasing $\dot{\gamma}_0^\mathrm{OSP}/\dot{\gamma}_0$ is accompanied by a reduction in particle contacts. To verify this, we calculate the time-averaged per-particle coordination number $\bar{Z}$. We find that $\bar{Z}(\gamma^\mathrm{OSP}_0, \omega)$ collapses in a similar fashion to the measured viscosity, and starts to decay around $\dot \gamma^\mathrm{OSP}_0 / \dot \gamma_0 \approx 1$, Fig.~\ref{figure2}(e). Indeed, plotting $\eta_\parallel$ versus $\bar{Z}$ we recover a simple relationship $\ln \eta \propto \bar{Z} + \mathrm{const}$, Fig.~\ref{figure2}(f). When no particle contact is formed, $\eta_\parallel$ corresponds to the viscosity purely arising from the interparticle hydrodynamic interactions~\cite{cheng2002nature}. As $\bar{Z}$ increases, $\eta_\parallel$ rises and brings more particles into contact, which in turn produces a higher viscosity. Overall, these simulations reaffirm the bulk rheological effect of the orthogonal perturbation and reveal the microstructural mechanism by which the OSP flow can manipulate the force chains to ultimately eradicate the frictional contacts responsible for shear thickening.

\begin{figure} [h]
\centering
\includegraphics[width=0.37 \textwidth]{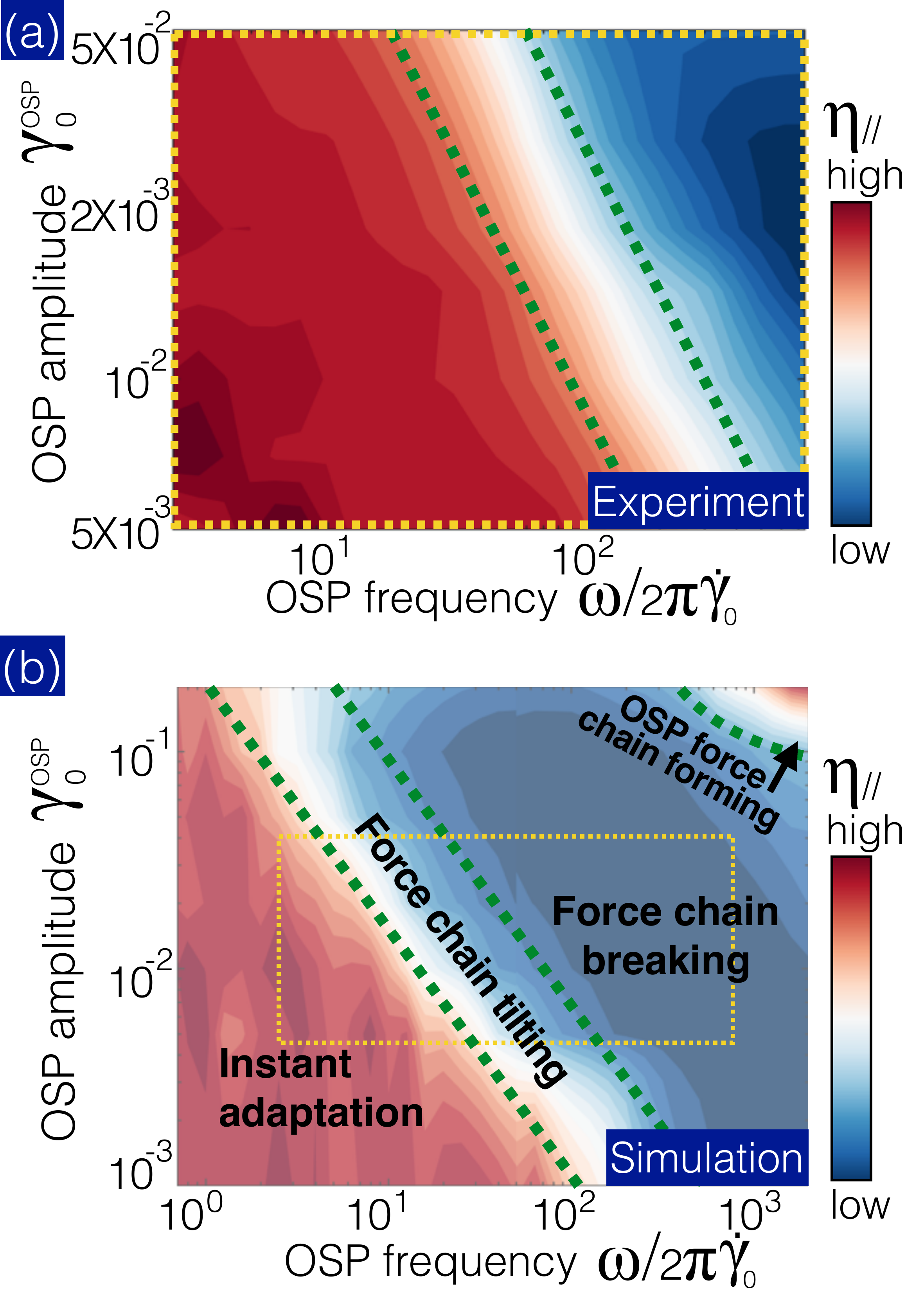}
\caption{Regime maps for strategic suspension viscosity regulation. Data are obtained from (a) experiments and (b) simulations. In (b), we delineate each of the phenomenological shear thickening regimes. Color scale represents primary flow viscosities. The yellow dashed box indicates the  regime explored in experiment. In both experiment and simulation, the onset of viscosity reduction occurs at $\dot \gamma^\mathrm{OSP}_0/\dot \gamma_0 \approx 1$. This indicates that the state of the biaxially sheared suspension is determined by the competition between primary and OSP flow rates.}
\label{figure3}
\end{figure}

We construct a summary phase diagram recapitulating our strategy for tuning shear thickening, giving experimental and simulated primary viscosities as functions of $\gamma^\mathrm{OSP}_0$ and $\omega$ in Fig.~\ref{figure3}(a) and (b), respectively. When an OSP flow is applied, its shear rate $\gamma_0^{\mathrm{OSP}} \omega$ determines the force chain behavior and the state of thickening in the suspension. The force chain behavior shows a transition from instant adaptation, chain tilting, to chain breaking as the OSP flow rate increases. When the OSP flow rate is even higher, however, the OSP flow may start to induce force chains by itself for sufficiently large $\gamma_0^\mathrm{OSP}$. A reentrant thickening behavior is then observed, as indicated by the curved contours and red region in the upper right corners of Fig.~\ref{figure3}(a) and (b), respectively.

\begin{figure} [h]
\centering
\includegraphics[width=0.37 \textwidth]{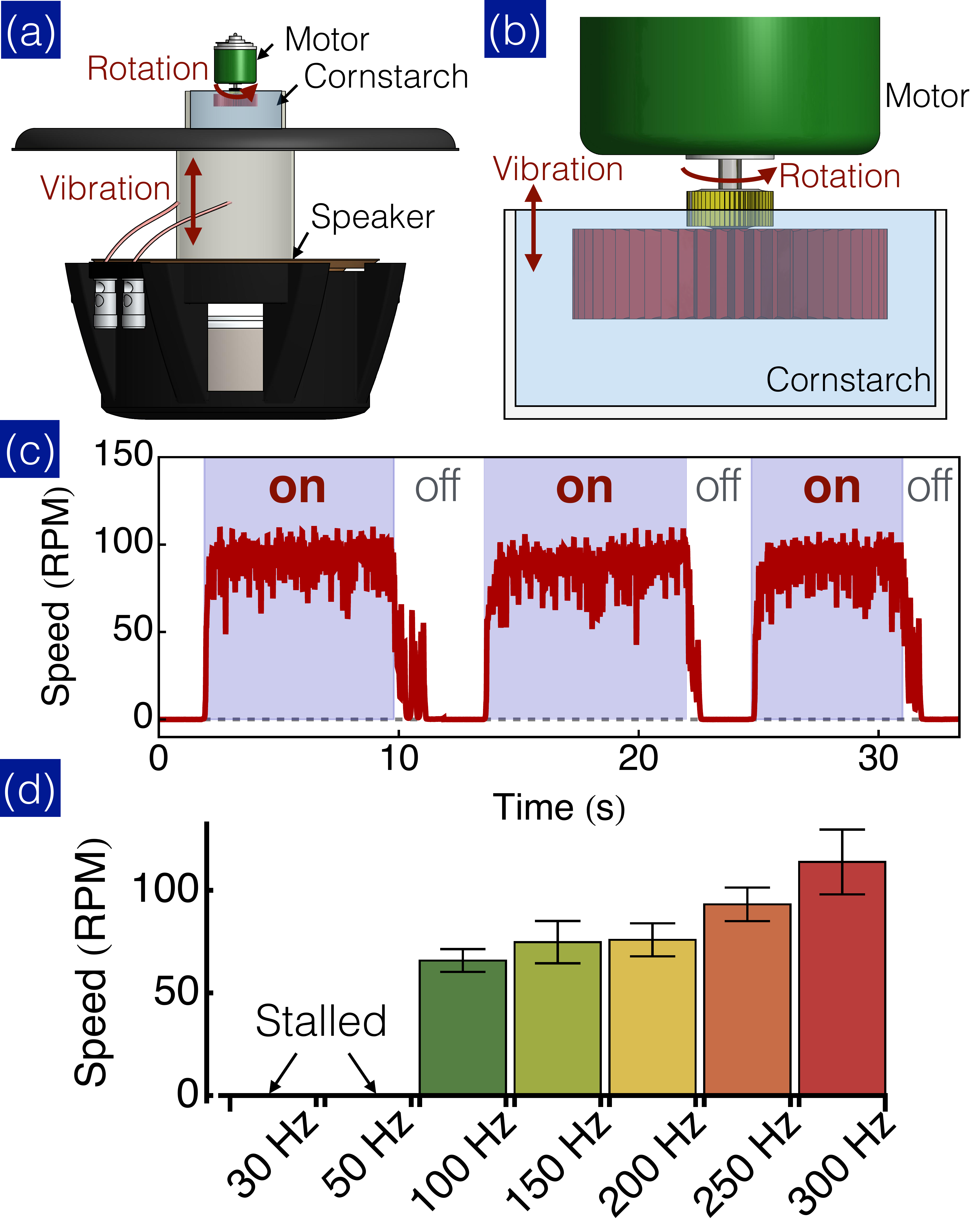}
\caption{A simple flow regulation set up comprising a rotating gear submerged in a cornstarch suspension illustrates the broad utility of our control scheme. Schematic of improvised flow regulator, showing (a) the full setup including driving motor, cornstarch reservoir and speaker and (b) a close-up of the setup. (c) The motor starts to rotate when the vibration (250 Hz) switches on (shaded areas), and stalls when the vibration is off. Shown in (d) is the rotation speed of the motor as a function of vibration frequency at a fixed speed. The motor remains stalled at the frequencies $\leq$ 50 Hz.}
\label{figure4}
\end{figure}

The orthogonal flow induced viscosity reduction allows us to drive the material at a higher speed with the same amount of driving force. In some practical cases, our biaxial protocol or a comparable oscillatory perturbation should be able to thin the suspension and effectively increase the flow processing speed. We test this prediction in a dense cornstarch suspension (solid fraction $\approx$ 40 wt \%)~\cite{fall2012shear, fall2008shear} using a simple setup consisting of a motor driven at a constant voltage and a speaker, as shown by the schematics in Fig.~\ref{figure4} (a) and (b). Upon motor startup, the rotating gear on the shaft immediately thickens the cornstarch, generating a large resistance that substantially slows down the gear speed. When the gear speed is below a threshold, the motor stalls. We then use the speaker to apply a high frequency and small amplitude (250Hz, 18 $\mu$m) vibration along the vertical direction to break the force chains against the gear. As a result, the motor under the same applied voltage starts to rotate again and drive the material, see Fig.4(c) and SI video. This transition repeats reversibly as we switch the speaker on and off. We also find that the motor rotation speed is maximized when we raise the vibration frequency and lower its amplitude, keeping the induced rate, analogous to $\dot{\gamma}_0^\mathrm{OSP}$, fixed Fig.~\ref{figure4}(d). This finding suggests that the state of the system is in the dark blue channel in the phase diagram (Fig.~\ref{figure3}). This channel runs upper left to bottom right indicating larger viscosity reductions at higher frequencies when the shear rate is fixed. Finally, we replace the motor with a stress-controlled rheometer (TA-Instrument DHR-3), which can rotate at a much slower rate without stalling, and repeat the speaker experiment (see SI). In this modified demonstration, the impeller on the rheometer rotates slowly when the vibration is off, and speeds up as the vibration is turned on. We can control such a transition between slow and fast rotations by simply switching the speaker vibration. These simple demonstrations show that our strategy for using orthogonal flows to control thickening is robust, and can be realized even in less controlled settings.

In conclusion, we have, for the first time, employed a biaxial shear protocol to tune the shear thickening viscosity of a suspension. Such a control is achieved by applying a transverse perturbation that regulates the main flow, along the lines of a mechanical transistor. This contrasts with passive control in which the rheological response is set when formulating the suspension and not changed thereafter. By scaling the flow measurement data and numerically investigating the force chain behavior, we elucidate the underlying mechanism of our control method and demonstrate how it can be strategically utilized. This result might inform extension of a three-dimensional continuum description so far limited to the $\dot \gamma^\mathrm{OSP}_0/\dot \gamma_0\ll1$ regime~\cite{Ovarlez2010}. 
Using a simpler flow regulation set up, we finally demonstrate that the insight obtained here can inform practical strategies, e.g., to unclog blockages caused by thickening during paste extrusion~\cite{patil2006constitutive}, 3D printing suspensions~\cite{sun20133d}, and flow of carbon black in energy storage devices~\cite{fan2014polysulfide}, and, more generally, to control bistability in granulation~\cite{Cates2014} and jamming in hopper flow~\cite{damond1939apparatus}. In general, control over a fluid's rheological properties holds the promise for advancing actuation and motion controls~\cite{stanway1996applications, mavroidis20005}, which have applications ranging from controllable dampers~\cite{stanway1996applications}, robotic arms~\cite{stanway1996applications, tan2002simple}, to actuating orthoses~\cite{nikitczuk2005rehabilitative}.

\section{Method}

The experiment is performed using an ARES-G2 rheometer (TA Instrument) in conjunction with an orthogonal superimposed perturbation (OSP) module~\cite{vermant1997orthogonal, vermant1998orthogonal}. During the experiment, the parallel ($\eta_\parallel$) and orthogonal ($|\eta_\bot^*|$) viscosities of the sample are measured by strain gauges on the upper shaft.
The tested suspension comprises silica particles in dipropylene glycol (Sigma-Aldrich). The sample volume fraction $\phi \approx 0.59$ is determined by directly imaging the suspension structure with a confocal microscope. The suspended particles are binary (2$\mu$m/4$\mu$m diameters) with number ratio roughly 1:1, mitigating crystallization. Our sample shear thickens at a relatively low shear rate due to the high solvent viscosity $\eta_0=84$ m.Pa, minimizing the instrumental instability in the biaxial test. For further characterizations of the sample structure and rheological properties, see SI.

The experimental volume fraction of 59\% is chosen to represent an upper limit at which steady flow in the shear thickening regime is observed. Numerically, we are able to reproduce the uniaxial experimental rheology at 55\% volume fraction. In future work we may address this difference by improving volume fraction measurement~\cite{poon2012measuring} and characterising particle friction~\cite{Fernandez2013}, both of which remain interesting experimental challenges.

\section{Acknowledgements}
IC and NL were supported by NSF CBET-PMP Award No. 1232666 and continued support from NSF CBET-PMP Award No. 1509308. CN and JS acknowledge funding from the Engineering and Physical Sciences Research Council (EP/N025318/1). MEC is supported by the Royal Society and EPSRC Grant EP/J007404. This work also made use of the Cornell Center for Materials Research Shared Facilities which are supported through the NSF MRSEC program (DMR-1120296). IC and NL gratefully acknowledge the Weitz lab at Harvard for generous use of their rheometry facility. The authors would also like to thank Jeffrey Morris, Jan Vermant, Dan Blair, Robert Behringer, Brian Leahy, John Brady, Andrea Liu, Romain Mari, John Royer and Wilson Poon for helpful discussions.

\end{document}